\documentclass[]{aa}
\usepackage{txfonts}
\usepackage{graphicx}
\usepackage{natbib}
\bibpunct{(}{)}{;}{a}{,}{,}

\begin{document}


\title{Testing SNe Ia distance measurement methods with SN~2011fe}

\author{J. Vink\'o\inst{1,2}\and 
K. S\'arneczky\inst{3,4}\and
K. Tak\'ats\inst{1}\and 
G. H. Marion\inst{5}\and
T. Heged\"us\inst{6}\and
I. B. B{\'\i}r\'o\inst{6}\and
T. Borkovits\inst{6,3,4}\and
E. Szegedi-Elek\inst{3}\and
A. Farkas\inst{7,3}\and
P. Klagyivik\inst{3}\and
L. L. Kiss\inst{3,8}\and
T. Kov\'acs\inst{3}\and
A. P\'al\inst{3,7}\and
R. Szak\'ats\inst{3}\and
N. Szalai\inst{3}\and 
T. Szalai\inst{1}\and
K. Szatm\'ary\inst{9}\and
A. Szing\inst{3}\and
K. Vida\inst{3}\and
J. C. Wheeler\inst{2}
}

\institute{Department of Optics and Quantum Electronics, University of Szeged,
D\'om t\'er 9, Szeged, 6720 Hungary\\
\email{vinko@physx.u-szeged.hu}\and
Department of Astronomy, University of Texas at Austin, 1 University Station C1400, Austin, TX 78712-0259, USA\and
Konkoly Observatory, MTA CSFK, Konkoly Th. M. \'ut 15-17., 1121 Budapest, Hungary \and
ELTE Gothard-Lend\"ulet Research Group, 9700 Szombathely, Hungary \and 
Harvard-Smithsonian Center for Astrophysics, Garden St. 60, Cambridge, MA, USA \and
Baja Astronomical Observatory, POB 766, Baja, 6500 Hungary \and
Department of Astronomy, Lor\'and E\"otv\"os University, POB 32, Budapest, 1518 Hungary \and
Sydney Institute for Astronomy, School of Physics A28, University of Sydney, NSW 2006, Australia \and
Department of Experimental Physics, University of Szeged, D\'om t\'er 9, Szeged, 6720 Hungary
}

\date{Received ; accepted }

\abstract{}
{The nearby, bright, almost completely unreddened Type Ia supernova 2011fe in M101 provides a
unique opportunity  to test both the precision and the accuracy of the 
extragalactic distances derived from SNe Ia light curve fitters.}
{We apply the current, public versions of the independent light curve fitting
codes MLCS2k2 and SALT2 to compute the distance modulus of SN 2011fe from
high-precision, multi-color (BVRI) light curves.}
{The results from the two fitting codes confirm that 2011fe is a ``normal'' (not 
peculiar) and only slightly reddened SN Ia. New unreddened distance moduli are
derived as $29.21$ $\pm$ 0.07 mag ($D \sim 6.95~ \pm ~0.23$ Mpc, MLCS2k2), and 
$29.05 ~\pm 0.07$ mag ($6.46~ \pm ~0.21$ Mpc). }
{Despite the very good fitting quality achieved with both light curve fitters, the
resulting distance moduli are inconsistent by $2 \sigma$. 
Both are marginally consistent (at $\sim 1 \sigma$) with the HST
Key Project distance modulus for M101. The SALT2 distance is in good
agreement with the recently revised Cepheid- and TRGB-distance to M101 by
Shappee \& Stanek. Averaging all SN- and Cepheid-based estimates,
the absolute distance to M101 is $\sim 6.6 ~\pm~ 0.5$ Mpc.}


\titlerunning{Distance to M101 from SN~2011fe}
\authorrunning{Vink\'o et al.}

\maketitle

\section{Introduction}\label{intro}

Supernovae (SNe) Ia are extensively used for deriving extragalactic distances,
because in the last two decades they turned out to be precise and reliable 
distance indicators \citep[see e.g.][and references therein]{mathe12}.
Although they are {\it not} standard candles in the optical (contrary to the widespread statements
that they are, which appear frequently even in the most recent papers), their
light curve (LC) shape correlates with their peak absolute magnitude, making
them {\it standardizable} (or calibratable) objects. The main advantage of applying
SNe Ia for distance measurement is that the method is essentially photometric and does not
need spectroscopy, save for typing the SN as a Ia. 

The LC shape is connected with the peak brightness via the 
``Phillips-relation'' \citep{phil93}, which states that
SNe Ia that decline more slowly after maximum have intrinsically
brighter peak magnitude, and vice versa. Although there are attempts
to explain this phenomenon based on theoretical grounds \citep{hoef86, pe01, kw07},  
this ``peak magnitude -- LC properties'' relation is still mostly
empirical at present. Therefore, the whole procedure of getting distances from
the photometry of SNe Ia relies on empirical calibrations
of the SNe Ia peak brightnesses, which need accurate, independent distances
as well as other details like reddening and intrinsic color for the calibrating objects. 
This is the major source of the relatively small,
but still existing random and systematic errors that limit the precision
and accuracy of the derived distances \citep[see e.g.][for 
further discussion and references]{mandel11}. 

SN~2011fe \citep[aka PTF11kly,][]{nugent11} is an excellent object in this respect,
because this bright ($m_{peak} \sim 10$ mag), nearby ($D \sim 6.5$ Mpc)
SN Ia was discovered hours after explosion \citep{nugent11, bloom11} and
suffered from very little interstellar reddening 
\citep[$A_V \sim 0.04$ mag,][]{nugent11, patat11}. The high apparent brightness 
allowed us to obtain accurate, high signal-to-noise photometry, while the very low redshift
\citep[$z = 0.000804$,][]{rc3} of its host galaxy, M101 (NGC~5457), eliminates
the necessity of K-corrections for the photometry that otherwise would be
a major source of systematic errors plaguing the distance determination
\citep{h07, h09}. The low interstellar reddening is also a very fortunate
circumstance, because all complications regarding the handling of
the effect of interstellar dust (galactic vs. non-standard reddening,  
disentangling reddening and intrinsic color variation, etc.) are expected to
be minimal. Also, the host galaxy, M101, has many recently published
distance estimates by various methods including Cepheids \citep{kp1, ss11}.
Therefore, SN~2011fe is an ideal object to test the current state-of-the-art 
of the SN Ia LC fitters.

In this paper we present new, homogeneous, calibrated ({\it BVRI}) photometry for
SN~2011fe obtained with a single telescope/detector combination (Sect.~\ref{obs}). 
We apply the two most widely accepted and trusted, independently calibrated, public LC fitters for SNe Ia, 
{\tt MLCS2k2} \citep{jrk07} and {\tt SALT2} \citep{guy07} to derive photometric 
distances to M101 from our data (Sect.~\ref{ana}). The results are compared
with other M101 distance estimates in Sect.~\ref{disc}. Section~\ref{conc} summarizes our results. 

\section{Observations}\label{obs}

\subsection{Photometry of SN~2011fe}

\begin{table}
\centering
\caption{\label{tab-phot} {\it BVRI} magnitudes of SN~2011fe from the Konkoly
Observatory, Hungary}
\begin{tabular}{ccccc}
\hline
\hline
JD\tablefootmark{a} & B (mag) & V (mag) & R (mag) & I (mag)\\
\hline
799.3 & 14.394 (.05) & 14.079 (.05) & 13.982 (.05) & 13.985 (.05) \\
800.3 & 13.688 (.05) & 13.277 (.05) & 13.230 (.05) & 13.167 (.05) \\
801.3 & 12.880 (.04) & 12.603 (.04) & 12.580 (.04) & 12.514 (.04) \\
802.3 & 12.203 (.02) & 12.176 (.02) & 12.069 (.02) & 12.039 (.02) \\
803.3 & 11.738 (.02) & 11.761 (.02) & 11.656 (.02) & 11.607 (.02) \\
804.3 & 11.327 (.02) & 11.418 (.02) & 11.264 (.02) & 11.246 (.02) \\
805.3 & 11.054 (.04) & 11.051 (.04) & 10.958 (.04) & 10.953 (.04) \\
807.3 & 10.480 (.05) & 10.595 (.05) & 10.509 (.05) & 10.519 (.05) \\
808.3 & 10.299 (.03) & 10.479 (.03) & 10.372 (.03) & 10.412 (.03) \\
809.3 & 10.141 (.04) & 10.317 (.04) & 10.276 (.04) & 10.328 (.04) \\
811.3 & 9.955 (.14) & 10.179 (.14)  & 10.042 (.14) & 10.219 (.14) \\
815.3 & 10.027 (.06) & 10.094 (.06) & 10.108 (.06) & 10.309 (.06) \\
816.3 & 10.005 (.07) & 10.113 (.07) & 10.037 (.07) & 10.383 (.07) \\
817.2 & 10.078 (.08) & 10.081 (.08) & 10.080 (.08) & 10.370 (.08) \\
818.2 & 10.063 (.05) & 10.049 (.05) & 10.042 (.05) & 10.448 (.05) \\
819.2 & 10.136 (.07) & 10.078 (.07) & 10.104 (.07) & 10.525 (.07) \\
820.3 & 10.096 (.06) & 10.020 (.06) & 10.091 (.06) & 10.485 (.06) \\
821.3 & 10.211 (.05) & 9.995 (.05) & 10.080 (.05) & 10.493 (.05) \\
822.3 & 10.138 (.07) & 10.058 (.07) & 10.162 (.07) & 10.562 (.07) \\
826.3 & 10.575 (.06) & 10.330 (.06) & 10.487 (.06) & 10.879 (.06) \\
828.3 & 10.799 (.02) & 10.505 (.02) & 10.659 (.02) & 10.921 (.02) \\
829.3 & 10.956 (.02) & 10.566 (.02) & 10.730 (.02) & 10.918 (.02) \\
830.3 & 11.110 (.05) & 10.630 (.05) & 10.735 (.05) & 10.934 (.05) \\
831.3 & 11.181 (.03) & 10.654 (.03) & 10.717 (.03) & 10.873 (.03) \\
832.3 & 11.263 (.02) & 10.744 (.02) & 10.774 (.02) & 10.856 (.02) \\
835.2 & 11.689 (.02) & 10.909 (.02) & 10.763 (.02) & 10.785 (.02) \\
837.2 & 11.907 (.04) & 10.968 (.04) & 10.794 (.04) & 10.736 (.04) \\
839.2 & 12.076 (.06) & 11.009 (.06) & 10.795 (.06) & 10.602 (.06) \\
844.3 & 12.744 (.07) & 11.432 (.07) & 11.113 (.07) & 10.780 (.07) \\
849.2 & 12.961 (.02) & 11.762 (.02) & 11.395 (.02) & 11.060 (.02) \\
853.2 & 13.082 (.06) & 11.972 (.06) & 11.670 (.06) & 11.386 (.06) \\
856.6 & 13.253 (.21) & 12.085 (.21) & 11.805 (.21) & 11.579 (.21) \\
862.7 & 13.329 (.02) & 12.242 (.02) & 12.052 (.02) & 11.837 (.02) \\
867.2 & 13.430 (.02) & 12.452 (.02) & 12.230 (.02) & 12.067 (.02) \\
871.6 & 13.442 (.02) & 12.505 (.02) & 12.341 (.02) & 12.225 (.02) \\
872.7 & 13.451 (.02) & 12.571 (.02) & 12.374 (.02) & 12.305 (.02) \\
877.6 & 13.572 (.10) & 12.714 (.10) & 12.537 (.10) & 12.501 (.10) \\
881.7 & 13.605 (.08) & 12.840 (.08) & 12.661 (.08) & 12.680 (.08) \\
888.6 & 13.711 (.12) & 12.932 (.12) & 12.904 (.12) & 12.964 (.12) \\
889.6 & 13.691 (.05) & 13.032 (.05) & 12.932 (.05) & 13.027 (.05) \\
894.7 & 13.743 (.05) & 13.145 (.05) & 13.095 (.05) & 13.182 (.05) \\
\hline
\end{tabular}
\tablefoot{\tablefoottext{a}{JD - 2,455,000}. Errors are given in parentheses.}
\end{table}

\begin{table*}
\centering
\caption{\label{tab-std} Local tertiary standards in the field of SN~2011fe}
\begin{tabular}{ccccccc}
\hline
\hline
ID & RA (J2000) & Dec (J2000) & B (mag) & V (mag) & R (mag) & I (mag)\\
\hline
A &14:04:04.438 &+54:13:32.64 & 14.268 (0.028) &13.404 (0.010) &12.878 (0.019) &12.451 (0.015) \\
B &14:03:45.175 &+54:16:16.31 & 14.698 (0.033) &14.054 (0.014) &13.651 (0.025) &13.350 (0.022) \\
C &14:03:28.982 &+54:11:33.74 & 17.386 (0.127) &16.320 (0.056) &15.771 (0.090) &15.176 (0.082) \\
D &14:03:24.941 &+54:13:57.36 & 17.022 (0.117) &16.405 (0.059) &15.926 (0.096) &15.645 (0.091) \\
E &14:03:23.779 &+54:14:32.83 & 16.151 (0.068) &15.617 (0.034) &15.203 (0.056) &14.778 (0.052) \\
F &14:03:22.410 &+54:15:36.22 & 16.297 (0.055) &14.910 (0.022) &14.044 (0.036) &13.327 (0.032) \\
G &14:02:38.490 &+54:14:50.69 & 16.455 (0.085) &16.012 (0.044) &15.568 (0.072) &15.168 (0.068) \\
H &14:03:05.865 &+54:17:25.49 & 16.899 (0.101) &16.160 (0.049) &15.724 (0.081) &15.369 (0.075) \\
I &14:03:05.803 &+54:15:19.91 & 17.805 (0.162) &16.625 (0.069) &16.536 (0.122) &16.096 (0.111) \\
J &14:02:54.159 &+54:16:29.17 & 14.622 (0.032) &14.043 (0.013) &13.675 (0.024) &13.324 (0.021) \\
\hline
\end{tabular}
\tablefoot{Magnitude errors are given in parentheses.}
\end{table*}

We have obtained multi-color ground-based photometric observations for SN~2011fe
from the Piszk\'estet{\H o} Mountain Station of the Konkoly Observatory, Hungary.
We used the 60/90 cm Schmidt-telescope with the attached $4096 \times 4096$ 
CCD (FoV 70x70 arcmin$^2$, equipped with Bessel $BVRI$ filters). 
In Table~\ref{tab-phot} the data for the first 41 nights are presented.
Note that we use JD instead of MJD throughout this paper. 

\begin{figure}
\centering
\resizebox{\hsize}{!}{\includegraphics{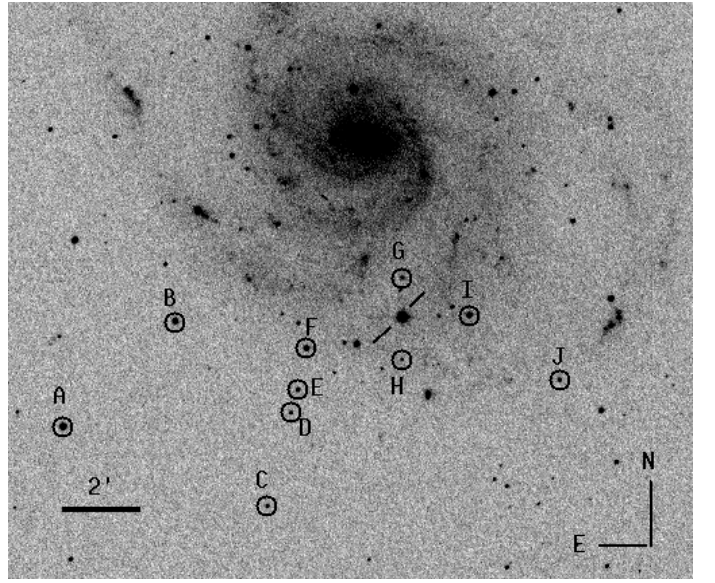}}
\caption{The field around SN 2011fe (marked by two line segments) with the local comparison stars encircled.}
\label{fig-schmidt}
\end{figure}

The magnitudes were obtained by applying aperture photometry using the 
{\it daophot/phot} task in 
$IRAF$\footnote{IRAF is distributed by the National Optical Astronomy
Observatories,
which are operated by the Association of Universities for Research
in Astronomy, Inc., under cooperative agreement with the National
Science Foundation.}. 
Because the background level around the SN position is relatively
low and uniform (see Fig.~\ref{fig-schmidt}), 
neither PSF-photometry, nor image subtraction were necessary to get reliable
light curves for SN~2011fe.
   
Transformation to the standard system was computed by using color terms 
expressed in the following forms for the $V$ magnitude and the
color indices:
\begin{eqnarray}
V-v~=~ C_V \cdot (V-I) + \zeta_{V} \nonumber \\
(B-V) ~=~ C_{BV} \cdot (b-v) + \zeta_{BV} \nonumber \\
(V-R) ~=~ C_{VR} \cdot (v-r) + \zeta_{VR} \nonumber \\
(V-I) ~=~ C_{VI} \cdot (v-i) + \zeta_{VI} ,
\label{eq1}
\end{eqnarray}
where lowercase symbols denote the instrumental magnitudes, while uppercase letters mean standard magnitudes.
The color terms were determined
by measuring Landolt standard stars in the field of PG2213 observed during photometric conditions:
$C_V = -0.019$, $C_{BV} = 1.218$, $C_{VR} = 1.035$, $C_{VI} = 0.959$.
These values were kept fixed while computing the standard magnitudes for the whole dataset.

Zero-points ($\zeta_{X}$) for each night were measured using local tertiary standard stars (Table~\ref{tab-std}).
These local comparison stars were tied to the Landolt standards during the photometric calibration.  

\begin{table}
\centering
\caption{\label{tab-baja} Unfiltered (scaled to $R$-band) photometry of SN~2011fe from the Baja Observatory, Hungary}
\begin{tabular}{cccc}
\hline
\hline
JD & R (mag) & J.D. & R (mag) \\
2455796.4 & $>$18.50 (0.48) & 2455822.3 & 10.21 (0.02)\\
2455800.3 & 13.11 (0.03) & 2455830.2 & 10.77 (0.02)\\
2455802.3 & 12.03 (0.02) & 2455831.3 & 10.81 (0.03)\\
2455804.3 & 11.25 (0.02) & 2455832.3 & 10.81 (0.02)\\
2455805.3 & 10.96 (0.02) & 2455834.2 & 10.88 (0.02)\\
2455808.3 & 10.41 (0.03) & 2455835.3 & 10.89 (0.02)\\
2455809.3 & 10.27 (0.02) & 2455837.2 & 10.95 (0.03)\\
2455811.3 & 10.10 (0.02) & 2455838.3 & 10.97 (0.02)\\
2455815.3 & 10.10 (0.03) & 2455863.2 & 12.16 (0.02)\\
2455817.3 & 10.01 (0.02) & 2455866.2 & 12.35 (0.03)\\
2455818.3 & 10.11 (0.02) & 2455867.2 & 12.34 (0.03)\\
2455819.3 & 10.12 (0.02) & 2455868.2 & 12.32 (0.03)\\
2455820.3 & 10.16 (0.03) & 2455871.7 & 12.44 (0.02)\\
\hline
\end{tabular}
\tablefoot{Errors are given in parentheses.}
\end{table}

Additional unfiltered photometry has been carried out at the Baja Observatory
of B\'acs-Kiskun County, Baja, Hungary with the 50 cm automated BART-telescope
equipped with a $4096 \times 4096$ back-illuminated Apogee Ultra CCD (FoV $40
\times 40$ 
arcmin$^2$, the frames were taken with $2 \times 2$ binning). During the course
of
the Baja-Szeged-Supernova-Survey (BASSUS) the field of M101 was imaged with BART
on 
2011-08-22.9 UT, $\sim 2$ days before discovery. No object was detected at the
position of SN~2011fe brighter than $\sim 18.5$ $R$-band magnitude. After
discovery,
unfiltered photometric observations were taken on 25 nights between Aug. 26 and
Nov. 6, 2011 (Table~\ref{tab-baja}). These data were scaled to the properly calibrated $R$-band
observations from Konkoly Observatory and used only in constraining the
the moment of explosion and the time of maximum light.  

\begin{figure}
\centering
\resizebox{\hsize}{!}{\includegraphics{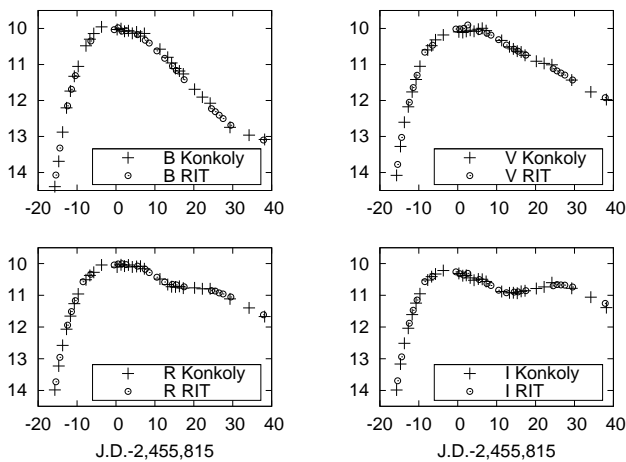}}
\caption{Comparison of LCs from Konkoly and RIT Observatories}
\label{fig-photcomp}
\end{figure}

Calibrated photometry for SN~2011fe have also been collected from 
recent literature. \citet{rich12} presented {\it BVRI} photometry 
obtained at the Rochester Institute of Technology (RIT) Observatory,
and at the Michigan State University Campus Observatory. A comparison
between the Konkoly and RIT data are plotted in Fig.~\ref{fig-photcomp}
(restricted to within 40 days around peak for better visibility)
illustrating the excellent agreement between these independent datasets.

\subsection{A UV-NIR spectrum of SN~2011fe}

\begin{figure}
\centering
\resizebox{\hsize}{!}{\includegraphics{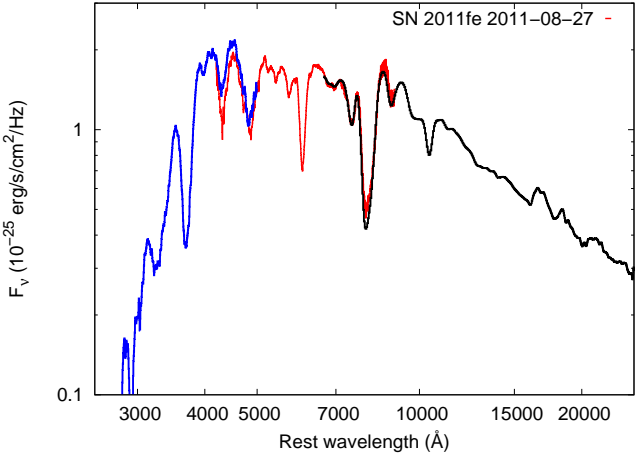}}
\caption{UV-NIR spectrum of SN 2011fe. Note that scales on both axes are logarithmic.}
\label{fig-sp}
\end{figure}

SN~2011fe was intensively followed up spectroscopically, and the
spectroscopic evolution is discussed in detail in \citet{parrent12} and 
\citet{smith12}. Expanding on that work is beyond the scope of this paper. 
Instead, we show only a single pre-maximum 
spectrum of SN~2011fe (Fig.~\ref{fig-sp}) extending from the ultraviolet (UV) to the 
near infrared (NIR). 

The spectrum plotted in Fig.~\ref{fig-sp} is a result of combining three datasets,
obtained with different instruments. The optical data were obtained with the 
HET Marcario Low Resolution Spectrograph 
(LRS, spectral coverage 4200 -- 10200 \AA, resolving power $\lambda / \Delta \lambda$
$\sim 600$) at the McDonald Observatory, Texas, on Aug. 27, 2011. These data were reduced
with standard IRAF routines. 
The UV part was taken by {\it Swift}/UVOT as a UGRISM observation on Aug. 28, 2011
\citep[see][]{brown12}. The reduction was done with the routine {\it uvotimgrism}
in HEASoft. The low ($R \approx 200$) and medium ($R \approx 1200$) resolution NIR 
spectra were obtained on August 26.3 UT with the 3.0 meter telescope at 
the NASA Infrared Telescope Facility (IRTF) using the SpeX 
medium-resolution spectrograph \citep{Rayner03}. The IRTF data were reduced 
using a package of IDL routines specifically designed for the reduction 
of SpeX data \citep[Spextool v. 3.4,][]{Cushing04}.

Fig.~\ref{fig-sp} illustrates the unprecedented quality of data available
for SN~2011fe, the analysis of which will be the topic of subsequent papers 
(e.g. a sequence of NIR spectra will be studied by Hsiao et al., in prep.).
A similar extended spectrum for the Type Ia SN~2011iv 
has been recently published by \citet{foley12}. The SN~2011fe spectrum presented
here is only the second such high-quality UVOIR spectrum for a Type Ia.
These kind of data may be especially useful for theoretical modeling.

Comparison with spectra of other SNe \citep{cenko11} revealed that SN~2011fe is a textbook-example of
Branch-normal SNe Ia. The spectroscopic evidence that SN~2011fe is a ``normal'' (i.e. not peculiar) SN Ia 
strengthens the applicability of the LC fitting techniques (see above) that were calibrated
for such SNe. 
  
\section{Analysis}\label{ana}

This section contains a brief review of the LC fitting methods for SNe Ia.
Their application to the observed data of SN~2011fe are then presented.

\subsection{SN Ia light curve fitters}

The empirical correlation between the LC shape and peak brightness of SNe~Ia 
was first revealed by \citet{phil93}, after the initial suggestion made by
\citet{psk77}. According to the Phillips-relation,
SNe Ia that decline more slowly after maximum are intrinsically brighter than
the more rapidly declining ones. \citet{phil93} introduced the $\Delta m_{15}(B)$
parameter for measuring the decline rate: it gives the decrease of the SN brightness
from the peak magnitude at 15 days after maximum in the $B$-band.  

This concept was further examined and extended to other photometric bands
by \citet{rpk}, introducing the ``Multi-Color Light Curve Shape'' ({\tt MLCS}) 
method. They defined a new parameter, $\Delta$, for measuring the peak
brightness as a function of the LC shape. Originally {\tt MLCS} was calibrated 
for the Johnson-Cousins $BVRI$ bands, and the LC in each band was described as
a linear combination of two empirical (tabulated) curves and the parameter $\Delta$.
These curves were calibrated (``trained'') using 9 nearby, well-observed SNe Ia that
had independent distances, mostly from the Tully-Fisher method. In \citet{riess98} {\tt MLCS}
was reformulated, expressing the LCs as a quadratic function of $\Delta$ and including
the $U$-band \citep[but see e.g. ][for discussion on the utility of the $U$-band data]{kessler09}. 

In this paper we applied the latest version, {\tt MLCS2k2} \citep{jrk07}, which
further improved the calibration by applying a sample of 133 SNe Ia for training
and also included a new parametrization for taking into account the effect of
interstellar reddening. In this version the observed LC
of a SN Ia can be expressed as
\begin{eqnarray}
m_x (t-t_0) ~=~ M_x^0 (t-t_0) + \mu_0 + \zeta_x (\alpha_x + {\beta_x \over R_V}) A_V^0 + \nonumber \\ 
~ ~ ~ + P_x (t-t_0) \cdot \Delta + Q_x (t-t_0) \cdot \Delta^2 ,
\label{eq2}
\end{eqnarray}
where $t-t_0$ is the SN phase in days, $t_0$ is the moment of maximum light in the $B$-band,
$m_x$ is the observed magnitude in the $x$-band ($x = B,V,R,I$), $M_x^0 (t-t_0)$ is the fiducial 
Ia absolute LC in the same band, $\mu_0$ is the true (reddening-free) SN distance modulus, 
$\zeta_x$, $\alpha_x$ and $\beta_x$ are functions describing the 
interstellar reddening, $R_V$ and $A_V^0$ are the ratio of total-to-selective absorption and the
$V$-band extinction at maximum light, respectively, $\Delta$ is the main LC parameter, and
$P_x$ and $Q_x$ are tabulated functions of the SN phase (``LC-vectors''). Together with $\Delta$, 
the functions $M_x^0$, $P_x$ and $Q_x$ describe the shape of the LC of a particular SN. 
For these functions we have applied the latest calibration downloaded from the {\tt MLCS2k2} 
website\footnote{\tt http://www.physics.rutgers.edu/\~{}saurabh/mlcs2k2/}. Note that 
\citet{jrk07} tied the {\tt MLCS2k2} LC-vectors to SNe Ia in the Hubble-flow adopting
$H_0 = 65$ kms$^{-1}$Mpc$^{-1}$. Thus, if needed, the distances
given by {\tt MLCS2k2} may be rescaled to other values of $H_0$ by
\begin{equation}
\mu_0(H_0) ~=~ \mu_0(\mathrm{MLCS}) - 5 \log_{10}(H_0 / 65) ~\mathrm{mag}.
\label{eq3}
\end{equation}
 
Another independent LC fitter, {\tt SALT2}, was developed by the
SuperNova Legacy Survey team \citep{guy07}. {\tt SALT2} is 
different from {\tt MLCS2k2} because it models the
whole spectral energy distribution (SED) of a SN Ia as
\begin{equation}
F_\lambda (p) ~=~ x_0 \cdot [ M_0(p,\lambda) + x_1 M_1(p, \lambda) ] \exp[c \cdot CL(\lambda)],
\label{eq4}
\end{equation}
where $p = t-t_0$ is the time from $B$-maximum (the SN phase), $F_\lambda$ is the phase-dependent
rest-frame flux density, $M_0(p,\lambda)$, $M_1(p,\lambda)$ and $CL(\lambda)$ are the {\tt SALT2} 
trained vectors. The free parameters $x_0$, $x_1$ and $c$ are the normalization- ,
stretch- and color parameters, respectively. We applied version 2.2.2b of
the code\footnote{\tt http://supernovae.in2p3.fr/\~{}guy/salt/usage.html}, which was trained with
the SNLS 3-year data \citep[][G10 hereafter]{guy10}. 

Because {\tt SALT2} models the entire SED, the observed LCs made with a particular filter set
must be derived by synthetic photometry. {\tt SALT2} performs this computation based on the
information provided by the user on the magnitude system in which the input data were taken. 
Since our photometry is in the Johnson-Cousins system (see Sect.~\ref{obs}),
we have selected the standard Vega-magnitude system given in the code.

\subsection{Constraining the moment of explosion and $B$-band maximum}\label{texp}

\begin{figure}
\centering
\resizebox{\hsize}{!}{\includegraphics{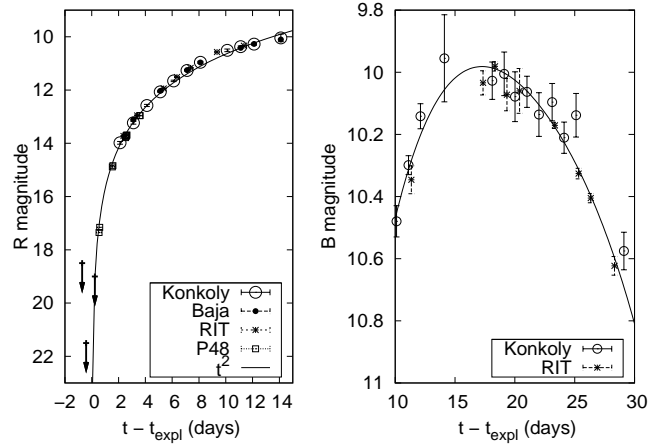}}
\caption{Fitting for the moment of explosion (left panel) and the time of $B$-band maximum (right panel)}
\label{fig-rise}
\end{figure}

Although the LC fitters applied in this paper use the time of $B$-band maximum light
as the zero point of the time, the moment of explosion is also a very important physical
parameter for SNe Ia. This can be inferred from the pre-maximum photometry. 
In this section we repeat the analysis of \citet{nugent11} to estimate this parameter
for SN~2011fe using more data and a better sampled early LC. 

The pre-maximum LC of SNe Ia can be surprisingly well described by the 
constant temperature ``fireball'' model \citep[e.g.][]{arnett82, nugent11, foley12a}.
In this simple model the adiabatic loss of the ejecta internal energy is just
compensated by the energy input from the radioactive decay of the $^{56}$Ni and
$^{56}$Co synthesized during the explosion. This results in a nearly constant
effective temperature and a luminosity governed only by the change of the photospheric
radius, which, assuming homologous expansion, can be approximated as $R_{ph} \sim t$,
giving $L \sim t^2$. Although this simple picture does not capture all the details
in the pre-maximum ejecta, it provides a surprisingly good fit to the observations
\citep[see also][]{riess99, hayden10, gane11}.

We have applied the function of $m = -2.5 \log_{10} (k \cdot (t - t_{exp})^2)$ to the observed pre-maximum $R$-band magnitudes
in Table~\ref{tab-phot} and \ref{tab-baja}, supplemented by similar data collected
from \citet{nugent11} and \citet{rich12}. The fit parameters were $k$ and $t_{exp}$, where the
latter was further constrained by the epochs of published non-detections \citep[][and Table~\ref{tab-baja}]{nugent11, bloom11}.  
The left panel of Fig.~\ref{fig-rise} shows the excellent agreement between the
observed data and the fit $t^2$ law. The fitting resulted in $t_{exp} = $JD $2,455,797.216 \pm 0.010$,
in perfect agreement with the value JD $2,455,797.187 ~\pm~ 0.014$ reported by \citet{nugent11}. 
Relaxing the $t^2$ constraint to $t^n$ and optimizing $n$ gave $n = 2.050 ~\pm~ 0.025$ and
$t_{exp} =$ JD $2,455,797.182 ~\pm~ 0.021$, which do not differ significantly from the results assuming 
$t^2$. We conclude that the explosion of SN~2011fe occured at JD $2,455,797.20 ~\pm~ 0.16$ 
(2011-08-23 16:48 UT $\pm$ 12 min).  

The $B$-band data from Table~\ref{tab-phot} and from \citet{rich12} are also used to constrain the
moment of $B$-maximum. Fitting a 4th order polynomial to the magnitudes obtained between +9 and +30
days after explosion resulted in $t_{Bmax} =$ JD $2,455,814.4 ~\pm~ 0.6$, which was used as
input for the LC fitter codes (see below). Thus, the $B$-band maximum occured $\sim 17.2$ days
after explosion, very similar to the value derived by \citet{foley12a} for SN~2009ig (17.13 days),
which was also discovered in less than a day after explosion. The average value for the majority
of ``normal'' SNe Ia is $\sim 17.4 ~\pm~ 0.2$ days \citep{hayden10}. This supports the conclusion from
spectroscopy that SN~2011fe is a ``normal'' SN Ia.

\subsection{Distance measurement}

We have applied both {\tt MLCS2k2} and {\tt SALT2} to the observed $BVRI$ data of SN~2011fe shown
in Sec.2.1. Before fitting, all data have been corrected for Milky Way reddening adopting 
$A_V = 0.028$ mag and $E(B-V) = 0.009$ mag \citep{sfd}. Note that these values are consistent
with the recent recalibration of Milky Way reddening by \citet{sf11}. 
 
Because of the low redshift of the host galaxy ($z = 0.000804$, see Sect.~\ref{intro}), $K$-corrections
for transforming the observed magnitudes to rest-frame bandpasses are negligible, thus, they were
ignored. We have not included $U$-band LC data \citep{brown12} in either fitting,
thus avoiding the persistent systematic uncertainties in modeling SNe Ia $U$-band data 
\citep[e.g.][]{kessler09}. 

Note that the absolute magnitudes of SNe Ia LCs were calibrated to different peak magnitudes in
the two independent LC-fitters. {\tt MLCS2k2} was tied to SNe Ia distances assuming $H_0 = 65$
km s$^{-1}$Mpc$^{-1}$ \citep{jrk07}, while the peak magnitude in {\tt SALT2} was fixed assuming
$H_0 = 70$ km s$^{-1}$Mpc$^{-1}$ (G10). To get rid of this discrepancy, we have
transformed all distance moduli given below to $H_0 = 73$ km s$^{-1}$Mpc$^{-1}$ using Eq.~\ref{eq3}. 

Since the {\tt MLCS2k2} templates are defined between $-10$ and $+90$ days around
$B$-maximum, while the {\tt SALT2} templates extend from $-20$ to only $+50$ days,  
we performed the LC fitting not only for all observed data (listed in Table~\ref{tab-phot}),
but also for those obtained between $-7$ and $+40$ days around $B$-maximum
(JD $2,455,808 < t < 2,455,855$). This test was performed
in order to reach maximum compatibility between the applications of the two methods, 
and reduce the systematics that might bias the fitting results.

\subsubsection{MLCS2k2}

\begin{figure}
\centering
\resizebox{\hsize}{!}{\includegraphics{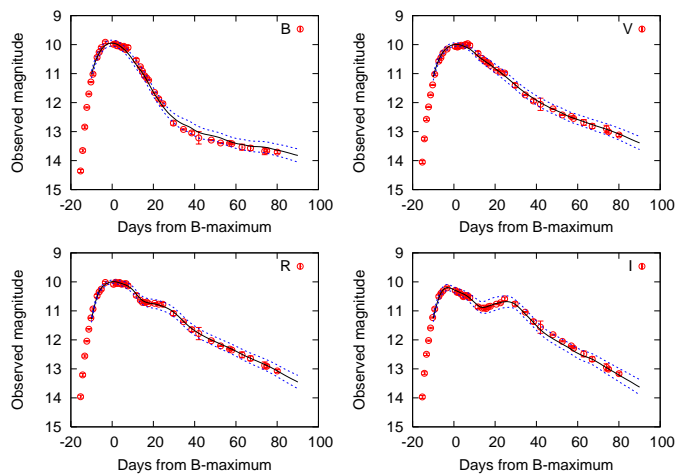}}
\caption{The fitting of the {\tt MLCS2k2} LCs to all observed data of SN~2011fe. Solid curves 
represent the best-fitting templates, while dotted curves denote the template uncertainties
given by the time-dependent variance of each template curve.}
\label{fig-mlcs2k2}
\end{figure}

\begin{table}
\centering
\caption{\label{tab-mlcs2k2} {\tt MLCS2k2} best-fitting parameters}
\begin{tabular}{lll}
\hline
\hline
Parameter & All data &  $-7 \mathrm{d} < t < + 40 \mathrm{d}$ \\
\hline
$t_0$ (JD) & 2,455,814.60 (0.10) & 2,455,814.9 (0.20) \\ 
$\Delta$ (mag) &  $-0.01$ (0.08) &  0.02 (0.08) \\
$A_V^{host}$ (mag) & 0.05 (0.01) &  0.00 (0.02) \\ 
$\mu_0(H_0=\mathrm{73})$ (mag) & 29.21 (0.07) & 29.23 (0.07) \\
Reduced $\chi^2$ & 0.2682 & 0.2507\\
\hline
\end{tabular}
\tablefoot{Errors are given in parentheses.}
\end{table}

\begin{figure}
\centering
\resizebox{\hsize}{!}{\includegraphics{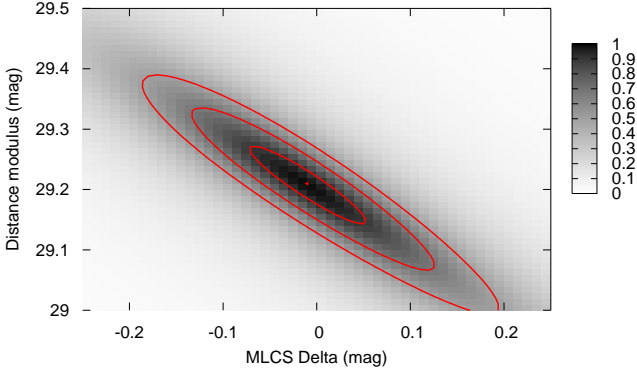}}
\caption{The map of the $\chi^2$ hypersurface around the minimum for {\tt MLCS2k2} fitting 
to all data. 
The contours correspond to 68, 90 and 95 \% confidence intervals (from inside to outside), respectively.}
\label{fig-chi2}
\end{figure}

The fitting of Eq.~\ref{eq2} was performed by using a simple, self-developed $\chi^2$-minimization code,
which scans through the allowed parameter space with a given step and finds the lowest $\chi^2$ within
this range. The fit parameters were the moment of $B$-maximum ($t_0$), the $V$-band extinction
$A_V$, the LC-parameter $\Delta$ and the distance modulus $\mu_0$, with steps of $\delta t_0 = 0.1$, 
$\delta A_V = 0.01$, $\delta \Delta = 0.01$ and $\delta \mu_0 = 0.01$, respectively. At the expense of
longer computation time, this approach maps the entire $\chi^2$ hypersurface and finds the absolute
minimum in the given parameter volume. 

We have fixed the reddening-law parameter as $R_V = 3.1$ appropriate for Milky Way reddening, although 
several recent results suggest that some high-velocity SNe Ia can be better modeled with significantly 
lower $R_V$ \citep{wang09, fk11}. Since SN~2011fe suffered from only minor reddening and most of it
is due to Milky Way dust (see below), it is more appropriate to adopt the galactic reddening law.  
Nevertheless, because of the low reddening, the value of $R_V$ has negligible effect on the final distance. 

The best-fitting {\tt MLCS2k2} model LCs are plotted together with the data in Fig.~\ref{fig-mlcs2k2}
(we plot only the results of fitting the whole dataset, because the fit to the restricted data range
gave very similar results). 
The final parameters are given in Table~\ref{tab-mlcs2k2} for both the whole and the
restricted data. The $1 \sigma$ uncertainties were estimated from the
contour of $\Delta \chi^2 = 1$ corresponding to 68\% confidence interval. Fig.~\ref{fig-chi2} shows
the map of the the $\chi^2$ hypersurface and the shape of the contours around the minimum for the two key parameters
$\Delta$ and $\mu_0$. It is seen that $\mu_0$ is strongly correlated with $\Delta$, which is the major source
of the relatively large uncertainty $\delta \mu = 0.07$ mag, despite the very good fitting quality.  

As seen in Table~\ref{tab-mlcs2k2}, there is no significant difference between the fit parameters
for the two datasets. The host extinction ($A_V^{host}$) is slightly less in the case of the
restricted dataset, but that is compensated by the higher value of $\Delta$ (meaning
fainter peak brightness), resulting in almost the same distance modulus. Thus, in the followings 
we adopt the parameters from fitting the full observed LC (left column in Table~\ref{tab-mlcs2k2}) 
as the final result from the particular LC-fitter, since those are based on the maximum available
information.

Note that the best-fitting {\tt MLCS2k2} template LC corresponds to $\Delta m_{15}(B) = 1.07 ~\pm~ 0.06$.
Although \citet{rich12} gives $1.21 ~\pm~ 0.03$ for this parameter, this is probably a misprint since
we measured $1.12 ~\pm~ 0.05$ from their published data, similar to the value of $1.10 ~\pm~ 0.05$ 
given by \citet{tammann}. It seems that all these parameters are consistent with each other, 
as well as with the finding that SN~2011fe has a nearly perfect fiducial SN Ia LC.
 
\subsubsection{SALT2}

\begin{figure}
\centering
\resizebox{\hsize}{!}{\includegraphics{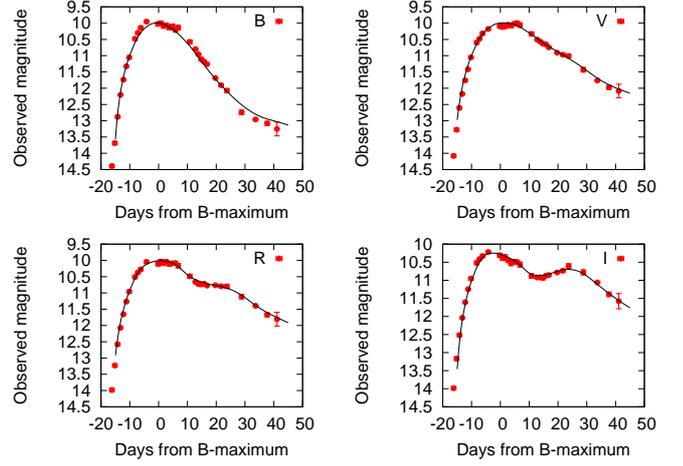}}
\caption{The fitting of {\tt SALT2} LCs to the observed data of SN~2011fe.}
\label{fig-salt2}
\end{figure}

The {\tt SALT2} fitting was computed by running the code as described on the
{\tt SALT2} website. The fit parameters provided by the code are:
$m_B^\star$ (rest-frame $B$-magnitude at maximum), $x_1$ (stretch) and $c$ (color). 
Note that {\tt SALT2} restricts the fitting to data obtained no later than
$+40$ days after maximum. 

Contrary to {\tt MLCS}, the {\tt SALT2} model does not explicitly include the distance or 
the distance modulus, thus, it must be derived from the fitting parameters. 
We followed two slightly different procedures for this: the one presented
by G10 and an independent realization given by \citet[][K09 hereafter]{kessler09}. 
Starting from the fitting parameters $m_B^\star$, $x_1$ and $c$, 
the distance modulus $\mu_0$ in the G10 calibration can be obtained as
\begin{eqnarray}
m_{BB} &=& m_B^\star - 0.008(\pm 0.005) \cdot x_1 + 0.013(\pm 0.004) \nonumber \\
CC &=& 0.997(\pm 0.097) \cdot C + 0.002(\pm 0.009) \cdot x_1 + 0.035(\pm 0.008) \nonumber \\
s &=& 0.107(\pm 0.006) \cdot x_1 + 0.991(\pm 0.006) \nonumber \\
\mu_0 &=& m_{BB} - M_B + a \cdot (s - 1) - b \cdot CC,
\label{eq5}
\end{eqnarray}
where we have adopted $M_B = -19.218 ~\pm~ 0.032$, $a = 1.295 ~\pm~ 0.112$ and $b=3.181 ~\pm~ 0.131$ 
(G10). 

The K09 calibration applies a simpler formula:
\begin{equation}
\mu_0 ~=~ m_B^\star - M_0 + \alpha \cdot x_1 - \beta \cdot c,
\label{eq6}
\end{equation}
where $M_0 = -19.157 ~\pm~ 0.025$, $\alpha = 0.121 ~\pm~ 0.027$ and $\beta = 2.63 ~\pm~ 0.22$ 
have been adopted from K09. 

The uncertainties in the formulae above were taken into account by a Monte-Carlo technique:
we calculated 10,000 different realizations of the above parameters by adding Gaussian
random numbers having standard deviations equal to the uncertainties above to the mean
values of all parameters and derived $\mu_0$ from each randomized set of parameters
applying Eq.~\ref{eq5} and \ref{eq6}. Then the average and the standard deviation of 
the resulting sample of $\mu_0$ values are adopted as the {\tt SALT2} estimate for the 
distance modulus and its uncertainty. Table~\ref{tab-salt2} lists the best-fitting 
parameters and errors, again for both the whole and the restricted dataset.
The final {\tt SALT2} distance modulus was obtained as
an unweigthed average of the two values from the G10 and K09 calibrations.

\begin{table}
\centering
\caption{\label{tab-salt2} {\tt SALT2} best-fitting parameters}
\begin{tabular}{lll}
\hline
\hline
Parameter & All data & $-7 \mathrm{d} < t < + 40 \mathrm{d}$ \\
\hline
$t_0$ (JD) & 2,455,815.505 (0.047) & 2,455,815.395 (0.097)\\ 
$m_B^\star$ (mag) & 9.959 (0.027) & 9.925 (0.029)\\
$x_1$ & $-0.296$ (0.049) & -0.360 (0.080)\\
$c$ & $-0.030$ (0.018) & -0.060 (0.020)\\ 
$\mu_0(\mathrm{G10})$ (mag) & 29.034 (0.078) & 29.088 (0.086)\\
$\mu_0(\mathrm{K09})$ (mag) & 29.068 (0.062) & 29.105 (0.068)\\
$\mu_0(\mathrm{final})$ (mag) & 29.05 (0.08) & 29.10 (0.08)\\
\hline
\end{tabular}
\tablefoot{Errors are given in parentheses. The final distance modulus (last row) 
is the average of the G10 and K09 estimates.}
\end{table}

\section{Discussion}\label{disc}

The application of {\tt MLCS2k2} and {\tt SALT2} LC-fitters for all observed data
resulted in distance moduli of $\mu_0$({\tt MLCS2k2}) = 29.21 $\pm$ 0.07 and 
$\mu_0$({\tt SALT2}) = 29.05 $\pm$ 0.08, respectively. It is seen that there is a $\sim 2 \sigma$
disagreement between these two values. Taking into account that
these distance moduli were obtained by fitting the same, homogeneous, densely-sampled,
high-quality photometric data of a nearby SN, and both codes provided
excellent fitting quality, this $\Delta \mu_0$ = 0.16 mag disagreement
is rather discouraging. Note that the difference exists despite correcting
the results from both codes to the same Hubble-constant, 
$H_0 = 73$ km s$^{-1}$Mpc$^{-1}$ (see above). 

Restricting the LC fitting only to data taken between $-7$ and $+40$ days 
around $B$-maximum (right column in Tables~\ref{tab-mlcs2k2} and \ref{tab-salt2}), 
the two distance moduli are both slightly higher and closer to one another: 
$\mu_0$({\tt MLCS2k2}) = 29.23 $\pm$ 0.07 and 
$\mu_0$({\tt SALT2}) = 29.10 $\pm$ 0.08, giving $\Delta \mu_0 = 0.13$ mag. 
Since these parameters are generally within the errors of those from fitting the
complete LC, the $\sim 2 \sigma$ disagreement still persists.

A similar, even larger difference of 
$\mu_0$({SALT2})$- \mu_0$({\tt MLCS2k2}) = $\pm$ 0.2 mag was found by K09
for the ``Nearby SNe'' sample of \citet{jrk07}, although deviations in both 
positive and negative directions have been revealed for individual SNe. 
Because the Nearby sample contains only a few very close, unreddened SNe
like SN~2011fe, the source of the mild discrepancy found by K09 is 
ambiguous. The present results suggest that, because of the lack of issues
due to reddening, $K$-correction, $U$-band anomaly or spectral peculiarities in the
case of SN~2011fe (see Sect.~\ref{intro}), the $>0.1$ mag difference between the {\tt MLCS2k2}
and {\tt SALT2} distance moduli is probably entirely due to a systematic offset between 
the different zero-point calibrations of the fiducial SN peak magnitude in the two methods. 

In order to test this statement,
we have compared the parameters in Table~\ref{tab-mlcs2k2} and \ref{tab-salt2}.
It is seen that for the whole dataset {\tt SALT2} estimates the $B$-maximum ($t_0$) as
being 0.9 day later than the epoch provided by {\tt MLCS2k2}. The result from the simple polynomial 
fitting (Sect.~\ref{texp}) is closer to the {\tt MLCS2k2} value, thus {\tt SALT2}
might tend to overestimate this parameter. For the restricted data the final 
$t_0$ from both methods changed slightly, becoming more similar, 
but a difference of $\sim 0.5$ day is still present. 

In order to investigate whether the uncertainty of $t_0$ could be responsible 
for the systematic difference between the distance moduli,
we have re-run the {\tt MLCS2k2} fitting for the restricted dataset 
by forcing $t_0$ equal to the {\tt SALT2} value of 2,455,815.395.
This resulted in $\Delta = 0.054$ mag and $\mu_0 = 29.205$ mag with $\chi^2 = 0.3242$ 
($A_V^{host}$ remained the same). The changes of the parameters are consistent with the 
shape of the $\chi^2$ surface plotted in Fig.~\ref{fig-chi2}: if $\Delta$ is increased, then $\mu_0$ 
decreases. This lower $\mu_0$ is indeed closer to the {\tt SALT2} value, but a 
systematic difference of $\sim 0.1$ mag still remains 
(i.e. the {\tt MLCS2k2} distance is still higher), while the quality of the fitting
is clearly worse. Thus, while forcing $t_0$ to be equal to the {\tt SALT2} value may 
somewhat reduce the systematic difference between the two methods, not all of the systematics 
affecting the distance modulus can be explained by this parameter alone. 

Distance measurements independent from these SN Ia LC-fitters may help in
resolving the open issue of absolute magnitude and distance calibrations.
Recently, \citet{mathe12} published distance estimates to SN~2011fe based
on near-infrared (NIR) photometry. Because SNe Ia appear to be much
better standard candles in the NIR than in optical bands, the usage
of good-quality NIR photometry for this bright, nearby SN looks promising.
Unfortunately, as \citet{mathe12} concluded, the present state-of-the-art
of getting SNe Ia NIR distances also suffers from an unsolved zero-point calibration problem.
This resulted in a wide range of NIR distance moduli for SN~2011fe spanning from 
28.84 to 29.14 mag (corrected to $H_0 = 73$, as above) with a mean value
of $\mu_0$(NIR) = 29.0 $\pm$ 0.2 mag \citep{mathe12}.  This is consistent
with the {\tt SALT2} result above, but it also agrees marginally  
(at $\sim 1 \sigma$) with the {\tt MLCS2k2} distance modulus. 
The range of the NIR distance moduli from different calibrations, 
0.31 mag \citep{mathe12}, is a factor of 2 larger 
than the uncertainties of the individual calibrations 
($\sim 0.15$ mag) of NIR peak magnitudes of SNe Ia. This may also be
a warning sign that the distance measurement technique from NIR LCs of SNe Ia 
is far from being settled.  

The situation is not much better if one considers the various distance estimates
available for the host galaxy, M101. This is one of the closest, brightest, and
most thoroughly studied galaxies for which Cepheid-based distances are available
\citep[see e.g.][and references therein]{mathe12}. The most widely accepted 
distance modulus of M101 is 29.13 $\pm$ 0.11 mag from the Cepheid $PL$-relation
by the $HST$ Key Project \citep{kp1}, which is just in the middle between 
the {\tt MLCS2k2} and {\tt SALT2} distance moduli above, being in $1 \sigma$
agreement with both. More recently \citet{ss11} obtained 29.04 $\pm$ 0.19
mag from an independent study of M101 Cepheids, which agrees better with the
{\tt SALT2} estimate, although its larger errors makes the result 
also consistent with {\tt MLCS2k2}. \citet{ss11} adopted the maser distance of
NGC~4258 as their distance anchor, which gives a lower distance modulus
for the LMC by 0.09 mag than the value adopted by \citet{kp1}. 
This accounts for most of the difference
between the two Cepheid-based results. Both of these Cepheid distances are slightly 
higher than the NIR-distance of \citet{mathe12}, but considering the larger errors
of the latter (0.2 mag), the three estimates are all more-or-less consistent with each other.

Non-Cepheid distance estimates to M101 span a $\sim 0.35$ mag wide range, from
29.05 \citep[from the Tip of the Red Giant Branch method, TRGB,][]{ss11} 
to 29.42 \citep[based on Planetary Nebulae Luminosity Function, PNLF,][]{feld96}, which does not
help much in resolving the issue of the M101 distance \citep[see Fig.~3 of ][]{mathe12}.

\section{Conclusions}\label{conc}

The nearby, bright, weakly reddened Type Ia supernova 2011fe 
in M101 provides a unique opportunity to test both the precision and the 
accuracy of the extragalactic distances derived from SNe Ia LC fitters. 
In this paper we presented new, calibrated {\it BVRI}-photometry for SN~2011fe. 
The LCs were analyzed with publicly available LC-fitters {\tt MLCS2k2} and {\tt SALT2}
to get the SN Ia-based distance to M101. There is a systematic offset of $\sim 0.15$ mag
between the {\tt MLCS2k2} and {\tt SALT2} distance moduli, the average of which also differs
by $\sim 0.13$ mag from the distance estimate by \citet{mathe12} from SN~2011fe NIR photometry.
This systematic offset between the results of the two widely-used LC-fitters may be
partly due to the different shape of the LC vectors near maximum, affecting the estimate
of the moment of maximum light, but the majority of the offset is probably caused
by systematic errors of the peak magnitudes from different photometric calibrations. 

We conclude that the weighted average of the three distance moduli of
SN~2011fe (using the inverse of the uncertainties as weights), 
$\mu_0$({\tt MLCS2k2}), $\mu_0$({\tt SALT2}) and $\mu_0$(NIR) \citep{mathe12}, 
and the two Cepheid-based distance to M101 \citep{kp1, ss11}
provides the following distance modulus of M101:
\begin{equation}
\mu_{0,\mathrm{M101}} ~=~ 29.109 ~\pm~ 0.049 ~\mathrm{(random)}~ \pm~ 0.1 ~\mathrm{(syst)} ~\mathrm{mag,} \nonumber
\end{equation}
which corresponds to $D_{\mathrm{M101}} = 6.6 ~\pm~ 0.5$ Mpc, taking into account both random and
systematic uncertainties. 
Despite the exceptional quality of the measured LCs of SN~2011fe
and the long history of efforts devoted to the calibration of LC-fitters for Ia SNe
as well as the absolute distance to M101, the available techniques cannot
predict the absolute distance to M101 with better than 0.5 Mpc ($\sim 8$\%) accuracy.

\begin{acknowledgements}

We are indebted to the referee, Peter Nugent, for his valuable comments and
suggestions that helped us improving the paper. 
This project has been supported by Hungarian OTKA grant K76816, by NSF Grant
AST 11-09881 (to JCW), and by the European Union together with the European Social Fund through 
the T\'AMOP 4.2.2/B-10/1-2010-0012 grant.
GHM and the CfA Supernova Program is supported by NSF Grant AST 09-07903.
GHM is a visiting Astronomer at the Infrared Telescope Facility, which 
is operated by the University of Hawaii under Cooperative Agreement no. 
NNX-08AE38A with the National Aeronautics and Space Administration, 
Science Mission Directorate, Planetary Astronomy Program.
AP has been supported by the ESA grant PECS~98073 and 
by the J\'anos Bolyai Research Scholarship of the Hungarian Academy of 
Sciences. TK has been supported by OTKA Grant K81373. 
KS, LLK and KV acknowledge support from the ''Lend\"ulet'' 
Young Researchers' Program of the Hungarian Academy of Sciences and the 
Hungarian OTKA Grants MB08C 81013, K83790 and K81421.
JCW is grateful for the hospitality of the Aspen Center for Physics
supported by NSF PHY-1066293. 
Thanks are also due to the staff of McDonald and Konkoly Observatories
for the prompt scheduling and helpful assistance during the time-critical
ToO observations. The Hobby-Eberly Telescope (HET) is a joint project of the 
University of Texas at Austin, the Pennsylvania State University, Stanford University, 
Ludwig-Maximilians-Universität München, and Georg-August-Universität Göttingen. 
The HET is named in honor of its principal benefactors, William P. Hobby and Robert E. Eberly.
The SIMBAD database at CDS, the NASA ADS and NED  databases have been used to access data and
references. The availability of these services is gratefully acknowledged.

\end{acknowledgements}

\bibliographystyle{aa} 
\bibliography{sn11fe_v3_aa.bbl} 
\end{document}